\documentclass[useAMS,usegraphicx,usenatbib]{mn2e}

\def\cqr{$\chi$$^2_\nu$}
\def\chisq{\chi^2}
\def\loglik{log(likelihood)}
\def\mloglik{--log(likelihood)}
\def\sext{$\sigma$$_{\rm ext}$}

\def\epo{$E_{\rm p,obs}$}
\def\epi{$E_{\rm p,i}$}
\def\omegam{$\Omega$$_{\rm M}$}
\def\omegal{$\Omega$$_{\Lambda}$}
\def\h0{H$_{\rm 0}$~}

\def\eiso{$E_{\rm iso}$}
\def\lpeiso{$L_{\rm p,iso}$}

\def\epeiso{$E_{\rm p,i}$ -- $E_{\rm iso}$}
\def\epega{$E_{\rm p,i}$ -- $E_{\gamma}$}
\def\nufnu{$\nu$F$_\nu$}

\def\sax{{\it Beppo}SAX}
\def\swift{{\it Swift}}
\def\flux{erg cm$^{-2}$ s$^{-1}$}

\def\ga{\mathrel{\hbox{\rlap{\hbox{\lower4pt\hbox{$\sim$}}}\hbox{$>$}}}}
\def\la{\mathrel{\hbox{\rlap{\hbox{\lower4pt\hbox{$\sim$}}}\hbox{$<$}}}}

\title[Cosmology with the \epeiso{} correlation in GRB]
{Measuring the cosmological parameters with the \epeiso{} correlation of Gamma--Ray 
Bursts} 
\author[L. Amati et al.]{Lorenzo Amati$^{1}$\thanks{E-mail:
amati@iasfbo.inaf.it}, Cristiano Guidorzi$^{2}$, Filippo 
Frontera$^{1,3}$,
Massimo Della Valle$^{4,5,6}$, \newauthor Fabio Finelli$^{1,7,8}$, Raffaella Landi$^{1}$
and Enrico Montanari$^{3,9}$
\\
$^{1}$INAF - IASF Bologna, 
via P. Gobetti 101, I-40129 Bologna (Italy)\\
$^{2}$INAF - Osservatorio Astronomico di Brera, via Bianchi, 46, I-23807 Merate 
(LC), Italy\\
$^{3}$Dipartimento di Fisica, Universit\`a di Ferrara, via Saragat 1, I-44100 
Ferrara, Italy\\
$^{4}$INAF - Osservatorio Astronomico di Capodimonte, Salita Moiariello, 16   
I-80131 Napoli (Italy)\\
$^{5}$European Southern Observatory -Karl Schwarschild Strasse 2, Garching bei Munchen
(Germany)\\
$^{6}$International Centre for Relativistic Astrophysics Network--Piazzale della Repubblica 2, Pescara (Italy)\\
$^{7}$INAF - Osservatorio Astronomico di Bologna, via Ranzani 1, I-40127 Bologna 
(Italy)\\  
$^{8}$INFN, Sezione di Bologna,
Via Irnerio 46, I-40126 Bologna (Italy)\\   
$^{9}$I.I.S. "I. Calvi", via Digione, 20, I-41034 Finale Emilia (MO), Italy
}
\begin{document}

\date{Submitted 2008 May 3.}

\pagerange{\pageref{firstpage}--\pageref{lastpage}} \pubyear{2008}

\maketitle

\label{firstpage}

\begin{abstract}
We have used the \epeiso{} correlation of GRBs to measure the
cosmological parameter \omegam. By adopting a maximum 
likelihood approach which
allows us to correctly quantify the extrinsic (i.e. non--Poissonian)
scatter of the correlation, we constrain (for a flat universe)
\omegam{} to 0.04--0.40 (68\% confidence level), with a best fit value
of \omegam{} $\sim 0.15$, and exclude \omegam{} = 1 at $>$ 99.9\%
confidence level. If we release the assumption of a flat universe, we
still find evidence for a low value of \omegam{} 
(0.04--0.50 at 68\%
confidence level) and a weak dependence of the dispersion of the
\epeiso{} correlation on \omegal{} (with an upper limit of \omegal{}
$\sim 1.15$ at 90\% confidence level).
Our approach makes no assumptions on the \epeiso{} correlation and it
does not use other calibrators to set the ``zero' point of the
relation, therefore our treatment of the data is not affected by
circularity and the results are independent of those derived via type
Ia SNe (or other cosmological probes). 
Unlike other multi-parameters correlations, our analysis
grounds on only two parameters, then including a larger number (a
factor $\sim 3$) of GRBs and being less affected by systematics.
Simulations based on realistic extrapolations of ongoing (and future)
GRB experiments (e.g., \swift, Konus--Wind, GLAST) show that: i) the
uncertainties on cosmological parameters can be significantly
decreased; ii) future data will allow us to get clues on the ``dark
energy'' evolution.
\end{abstract}


\section{Introduction}

Gamma--ray Bursts (GRBs) are the brightest cosmological sources in the
universe, with isotropic luminosities up to 10$^{54}$ \flux and a
redshift distribution extending at least up to $z$ $\sim$6.3 (e.g.,
Tagliaferri et al. 2005\nocite{Tagliaferri05}). 
Thus, these sources may be interesting for
cosmological studies, if one can use them to provide measurements of
the cosmological parameters independently of other methods, like
the CMB (e.g., De Bernardis et al. 2000\nocite{Debernardis00};
Spergel et al. 2003\nocite{Spergel03}; Dunkley et al. 2008
\nocite{Dunkley08}; Komatsu et al. 2008\nocite{Komatsu08}),
type Ia SN (e.g., Perlmutter et al. 1998; Perlmutter et al. 1999;
Riess et al. 1998; Riess et al. 2004; Astier et al. 2006\nocite{Perlmutter98,Perlmutter99, Riess98,
Riess04}), Baryon Acoustic Oscillations (BAO) (e.g., Eisenstein et al. 2005; Tegmark et al. 2006;
Percival et al. 2007\nocite{Percival07,
Eisenstein05,Tegmark06,Percival07}), galaxy clusters (e.g.,
Rosati, Borgani \& Norman 2002; Voit 2005\nocite{Rosati02,Voit05}).
However, GRBs are not standard
candles, given that their luminosities span several orders of
magnitude under the assumption of both isotropic and collimated
emission ( e.g., Ghirlanda et al. 2006\nocite{Ghirlanda06}). In the
recent years, several attempts to "standardize" GRB have been made,
mainly on the basis of correlations involving: a) a GRB intensity
indicator, like the isotropic--equivalent radiated energy (\eiso) or
the isotropic--equivalent peak luminosity (\lpeiso); b) the photon
energy at which the time averaged \nufnu{} spectrum peaks ("peak
energy", \epi{}); c) other observables, like the break time of the
afterglow light curve, t$_{\rm b}$, or the "high signal time scale"
T$_{0.45}$ \citep{Ghirlanda04,Firmani06}.
Finally, \citet{Schaefer07} used different correlations
(spectrum--energy, time lag -- luminosity, peak luminosity --
variability) to construct a GRB Hubble diagram and constrain the
cosmological parameters.

These analyses have provided useful constraints on \omegam{} and \omegal{}
consistent with those derived from type Ia SNe, see, e.g., 
\citet{Ghirlanda06} for a review. However, the use of these correlations for
cosmology is controversial. For example, because of the lack of low redshift
GRBs they cannot be directly
calibrated. On
the other hand the calibration of a spectrum--energy correlation using
SNe-Ia as calibrators \citep{Kodama08,Liang08} is an interesting
attempt to use GRBs as cosmological probes, but it may look suspicious 
that this method gives very similar results to those obtained via SNe.
Thus, particular care and sophisticated statistical methodologies have
to be adopted in order to avoid circularity problems when constructing
a GRB Hubble diagram.
In addition, recent analyses based on updated samples
of GRBs showed that the dispersion of three--parameter correlations
could be significantly larger than thought
before \citep{Campana07,Ghirlanda07,Rossi08}.

In this article we investigate the possibility of constraining the
cosmological parameters from the \epeiso{} correlation. 
This
correlation was initially discovered on a small sample of \sax{}
GRBs with known redshift \citep{Amati02} and confirmed afterwards by
Swift observations \citep{Amati06}. Although it was the first 
"spectrum--energy" correlation discovered for GRBs (and the most
firmly established one, to date) it was never used for cosmology
purposes, because of its significant "extrinsic" scatter (i.e., 
a scatter in excess 
to the "intrinsic" Poissonian fluctuations of the data). However,
the conspicuous increasing of GRB discoveries combined with the fact
that \epeiso{} correlation needs only two parameters that are directly
inferred from observations (this fact minimizes the effects of
systematics and increases the number of GRBs that can be used, by a
factor $\sim 3$) makes the use of this correlation an interesting tool
for cosmology. 

Our study was motivated by our finding that, in the assumption of a flat universe,
the $\chisq$ obtained by fitting the 
\epeiso{} correlation with a power-law varies with the value of 
\omegam{} assumed to compute \eiso{}, has its minimum value for 
\omegam$\sim$0.25, very close the value obtained with SNe-Ia approach.
In this paper we show indeed that a small fraction of the extrinsic scatter is
due to to the choice of cosmological parameters, thus allowing us
to constrain \omegam{} and, to
a smaller extent, \omegal.
\begin{table*}
\begin{minipage}{155mm}
\caption{Values of redshift, "bolometric" fluence and cosmological
rest--frame spectral peak energy, \epi{}=\epo{}$\times$(1+$z$),
of long GRBs and XRFs with 
firm estimates of both
$z$ and \epo (70 events) as of April, 11 2008. These are the values that we used to estimate
cosmological parameters. The table includes also the values of \eiso{} computed by
assuming \h0=70 km s$^{-1}$ Mpc$^{-1}$, \omegam{} = 0.3 and
\omegal{} = 0.7 . 
The uncertainties are at 1$\sigma$ significance. 
The "Instruments" column reports the name of the experiment(s), or of 
the satellite(s), that provided the estimates of spectral parameters and
fluence (GRO = CGRO/BATSE, SAX = \sax, HET = HETE--2, KW = Konus--Wind, SWI = \swift). 
The last column reports the references for the spectral
parameters used to derive the fluence and \epi{}. Redshift values were taken
from the GRB table by J. Greiner and references therein 
(http://www.mpe.mpg.de/~jcg/grbgen.html).
}
\begin{tabular}{llllllc}
\hline
 GRB  & $z$  & Fluence$^{\rm (a)}$ & \epi  & \eiso$^{\rm (b)}$  & Instruments &  Refs. for $^{\rm (c)}$  \\ 
        &  & (10$^{-5}$ erg cm$^{-2}$) &  (keV)     & (10$^{52}$ erg)  &  &  spectrum  \\
\hline
 970228   &  0.695  &  1.3$\pm$0.1 &  195$\pm$64               &  1.60$\pm$0.12             &  SAX         & (1)  \\
 970508   &  0.835   & 0.34$\pm$0.07 &  145$\pm$43               &  0.61$\pm$0.13             &  SAX         & (1)  \\
 970828   &  0.958   & 12.3$\pm$1.4 &  586$\pm$117              &  29$\pm$3                  &  GRO         & (1)  \\
 971214   &  3.42    & 0.87$\pm$0.11 &  685$\pm$133              &  21$\pm$3                  &  SAX         & (1)  \\
 980326   &  1.0     & 0.18$\pm$0.04 &  71$\pm36$                &  0.48$\pm0.09$             &  SAX         & (1)  \\ 
 980613   &  1.096   & 0.19$\pm$0.03 &  194$\pm$89               &  0.59$\pm$0.09             &  SAX         & (1)  \\
 980703   &  0.966   & 2.9$\pm$0.3 &  503$\pm$64               &  7.2$\pm$0.7               &  GRO         & (1)  \\
 990123   &  1.60    & 35.8$\pm$5.8 &  1724$\pm$466 &  229$\pm$37                &  SAX/GRO/KW & (1)  \\
 990506   &  1.30    & 21.7$\pm$2.2 &  677$\pm$156  &  94$\pm$9                &  GRO/KW     & (1)  \\
 990510   &  1.619   & 2.6$\pm$0.4 &  423$\pm$42               &  17$\pm$3                  &  SAX         & (1)  \\
 990705   &  0.842   & 9.8$\pm$1.4 &  459$\pm$139  &  18$\pm$3                  &  SAX/KW     & (1)  \\
 990712   &  0.434   & 1.4$\pm$0.3 &  93$\pm$15                &  0.67$\pm$0.13             &  SAX         & (1)  \\
 991208   &  0.706   & 17.2$\pm$1.4 &  313$\pm$31               &  22.3$\pm$1.8              &  KW         & (1)  \\
 991216   &  1.02    & 24.8$\pm$2.5 &  648$\pm$134  &  67$\pm$7                  &  GRO/KW     & (1)  \\
 000131   &  4.50    & 4.7$\pm$0.8 &  987$\pm$ 416 &  172$\pm$30                &  GRO/KW     & (1)  \\
 000210   &  0.846   & 8.0$\pm$0.9 &  753$\pm$26               &  14.9$\pm$1.6              &  KW         & (1)  \\
 000418   &  1.12    & 2.8$\pm$0.5 &  284$\pm$21               &  9.1$\pm$1.7              &  KW         & (1)  \\
 000911   &  1.06    & 23.0$\pm$4.7 &  1856$\pm$371 &  67$\pm$14                 &  KW         & (1)  \\
 000926   &  2.07    & 2.6$\pm$0.6 &  310$\pm$20               &  27.1$\pm$5.9              &  KW         & (1)  \\
 010222   &  1.48    & 14.6$\pm$1.5 &  766$\pm$30               &  81$\pm$9                 &  KW         & (1)  \\
 010921   &  0.450   & 1.8$\pm$0.2 &  129$\pm$26               &  0.95$\pm$0.10             &  HET         & (1)  \\
 011121   &  0.36    & 24.3$\pm$6.7 &  1060$\pm$265             &  7.8$\pm$2.1               &  SAX/KW     & (2)  \\
 011211   &  2.14    & 0.50$\pm$0.06 &  186$\pm$24               &  5.4$\pm$0.6               &  SAX         & (1)  \\
 020124   &  3.20    & 1.2$\pm$0.1 &  448$\pm$148  &  27$\pm$3                  &  HET/KW     & (1)  \\
 020405   &  0.69    & 8.4$\pm$0.7 &  354$\pm$10               &  10$\pm$0.9                &  SAX/KW         & (2)  \\
 020813   &  1.25    & 16.3$\pm$4.1 &  590$\pm$151  &  66$\pm$16     &  HET/KW     & (1)  \\
 020819B  &  0.410   & 1.6$\pm$0.4 &  70$\pm$21                &  0.68$\pm$0.17             &  HET         & (1)  \\
 020903   &  0.250   & 0.016$\pm$0.004 &  3.37$\pm$1.79            &  0.0024$\pm$0.0006         &  HET         & (1)  \\
 021004   &  2.30    & 0.27$\pm$0.04 &  266$\pm$117              &  3.3$\pm$0.4               &  HET         & (1)  \\
 021211   &  1.01    & 0.42$\pm$0.05 &  127$\pm$52   &  1.12$\pm$0.13              &  HET/KW     & (1)  \\
 030226   &  1.98    & 1.3$\pm$0.1 &  289$\pm$66               &  12.1$\pm$1.3                &  HET         & (1)  \\
 030323   &  3.37    & 0.12$\pm$0.04 &  270$\pm$113              &  2.8$\pm$0.9               &  HET         & (3)  \\
 030328   &  1.52    & 6.4$\pm$0.6 &  328$\pm$55               &  47$\pm$3                  &  HET/KW     & (1)  \\
 030329   &  0.17    & 21.5$\pm$3.8 &  100$\pm$23   &  1.5$\pm$0.3   &  HET/KW     & (1)  \\
 030429   &  2.65    & 0.14$\pm$0.02 &  128$\pm$26               &  2.16$\pm$0.26             &  HET         & (1)  \\
 030528   &  0.78    & 1.4$\pm$0.2 &  57$\pm$9.0               &  2.5$\pm$0.3               &  HET         & (3)  \\
 040912   &  1.563   & 0.21$\pm$0.06 &  44$\pm$33                &  1.3$\pm$0.3               &  HET         & (4)  \\
 040924   &  0.859   & 0.49$\pm$0.04 &  102$\pm$35   &  0.95$\pm$0.09              &  HET/KW     & (1)  \\
 041006   &  0.716   & 2.3$\pm$0.6 &  98$\pm$20                &  3.0$\pm$0.9               &  HET         & (1)  \\
 050318   &  1.44    & 0.42$\pm$0.03 &  115$\pm$25               &  2.20$\pm$0.16             &  SWI         & (1)  \\
 050401   &  2.90    & 1.9$\pm$0.4 &  467$\pm$110              &  35$\pm$7                  &  KW         & (1)  \\
 050416A  &  0.650   & 0.087$\pm$0.009 &  25.1$\pm$4.2             &  0.10$\pm$0.01             &  SWI         & (1)  \\
 050525A   &  0.606   & 2.6$\pm$0.5 &  127$\pm$10               &  2.50$\pm$0.43             &  SWI         & (1)  \\
 050603   &  2.821   & 3.5$\pm$0.2 &  1333$\pm$107             &  60$\pm$4                  &  KW         & (1)  \\
 050820   &  2.612   & 6.4$\pm$0.5 &  1325$\pm277$             &  97.4$\pm$7.8               &  KW         & (5)  \\
 050904   &  6.29    & 2.0$\pm$0.2 &  3178$\pm$1094            &  124$\pm$13                &  SWI/KW         & (6)  \\
 050922C  &  2.198   & 0.47$\pm$0.16 &  415$\pm$111              &  5.3$\pm$1.7               &  HET         & (1)  \\
 051022   &  0.80    & 32.6$\pm$3.1 &  754$\pm$258  &  54$\pm$5                  &  HET/KW     & (1)  \\
 051109A   &  2.346   & 0.51$\pm$0.05 &  539$\pm$200              &  6.5$\pm$0.7               &  KW         & (1)  \\
 060115   &  3.53    & 0.25$\pm$0.04 &  285$\pm$34               &  6.3$\pm$0.9               &  SWI         & (7)  \\
 060124   &  2.296   & 3.4$\pm$0.5 &  784$\pm$285              &  41$\pm$ 6                 &  KW         & (8)  \\
 060206   &  4.048   & 0.14$\pm$0.03 &  394$\pm$46               &  4.3$\pm$0.9               &  SWI         & (7)  \\
 060218   &  0.0331  & 2.2$\pm$0.1 &  4.9$\pm$0.3              &  0.0053$\pm$0.0003         &  SWI         & (9)  \\
 060418   &  1.489   & 2.3$\pm$0.5 &  572$\pm$143              &  13$\pm$3                  &  KW         & (10)  \\
\hline
\end{tabular}
\end{minipage}
\end{table*}
\begin{table*}
\begin{minipage}{155mm}
\begin{tabular}{llllllc}
\hline
 GRB  & $z$  &  Fluence$^{\rm (a)}$ & \epi  & \eiso$^{\rm (b)}$ & Instruments &  Refs. for $^{\rm (c)}$  \\ 
        &  & (10$^{-5}$ erg cm$^{-2}$) &  (keV)     & (10$^{52}$ erg)  &  &  spectrum  \\
\hline
 060526   &  3.21    & 0.12$\pm$0.06 &  105$\pm$21               &  2.6$\pm$0.3               &  SWI         & (11)  \\  
 060614   &  0.125   & 5.9$\pm$2.4 &  55$\pm$45                &  0.21$\pm$0.09             &  KW         & (12)  \\
 060707   &  3.425  &  0.23$\pm$0.04 &  279$\pm$28               &  5.4$\pm$1.0               &  SWI         & (7)  \\
 060814   &  0.84   &  3.8$\pm$0.4 &  473$\pm$155              &  7.0$\pm$0.7               &  KW         & (13)  \\
 060908   &  2.43   &  0.73$\pm$0.07 &  514$\pm$102              &  9.8$\pm$0.9              &  SWI         & (7)  \\
 060927   &  5.60   &  0.27$\pm$0.04 &  475$\pm$47               &  13.8$\pm$2.0              &  SWI         & (7)  \\
 061007   &  1.261  &  21.1$\pm$2.1 &  890$\pm$124              &  86$\pm$9                &  KW/SUZ         & (14)  \\
 061121   &  1.314  &  5.1$\pm$0.6 &  1289$\pm$153             &  22.5$\pm$2.6              &  KW/SUZ         & (15)  \\
 061126   &  1.1588 &  8.7$\pm$1.0 &  1337$\pm$410             &  30$\pm$3                  &  SWI/RHE       & (16)  \\
 070125   &  1.547  &  13.3$\pm$1.3 &  934$\pm$148              &  80.2$\pm$8.0              &  KW         & (17)  \\
 071010B   &  0.947  &  0.74$\pm$0.37 &  101$\pm$20               &  1.7$\pm$0.9                   &  KW         & (18)  \\
 071020   &  2.145  &  0.87$\pm$0.40 &  1013$\pm$160             &  9.5$\pm$4.3                  &  KW         & (19)  \\
 071117   &  1.331  &  0.89$\pm$0.21 &  647$\pm$226              &  4.1$\pm$0.9               &  KW         & (20)  \\  
 080319B  &  0.937  &  49.7$\pm$3.8 &  1261$\pm$65              &  114$\pm$9                &  KW         & (21)  \\
 080319C  &  1.95   &  1.5$\pm$0.3 &  906$\pm$272              &  14.1$\pm$2.8              &  KW         & (22)  \\
 080411   &  1.03   &  5.7$\pm$0.3 &  524$\pm$70               &  15.6$\pm$0.9              &  KW         & (23)  \\
\hline
\end{tabular}
\begin{list}{}{}
\item[]Notes. 
$^{\rm (a)}$ Bolometric fluence computed in the [$1/(1+z)$ -- $10000/(1+z)$] {\rm keV}
energy range.\\
$^{\rm (b)}$ Computed by assuming \h0=70 km s$^{-1}$ Mpc$^{-1}$, \omegam{} = 0.3 and
\omegal{} = 0.7 .\\
$^{\rm (c)}$References for the spectral parameters and for the values and
uncertainties of \epi{} and \eiso{}:
(1) \citet{Amati06} and references therein; (2) \citet{Ulanov05} and refined analysis of \sax data;
(3) \citet{Sakamoto05}; (4) \citet{Stratta06}; (5) \citet{Cenko06}; (6) \citet{Krimm06};
(7) \citet{Sakamoto08a}; (8) \citet{Golenetskii06a}; (9) \citet{Campana06};
(10) \citet{Golenetskii06b}; (11) \citet{Schaefer07} and references therein;
(12) \citet{Amati07} and references therein; (13) \citet{Golenetskii06c};
(14) \citet{Mundell07} and references therein; (15) \citet{Ghirlanda07} and references therein;
(16) \citet{Perley08}; (17) \citet{Golenetskii07a}; (18) \citet{Golenetskii07b}; 
(19) \citet{Golenetskii07c}; (20) \citet{Golenetskii07d}; 
(21) \citet{Golenetskii08a};
(22) \citet{Golenetskii08b}; (23) \citet{Golenetskii08c}.
\end{list}
\end{minipage}
\end{table*}
\begin{figure*} 
\centerline{\includegraphics[width=9.2cm]{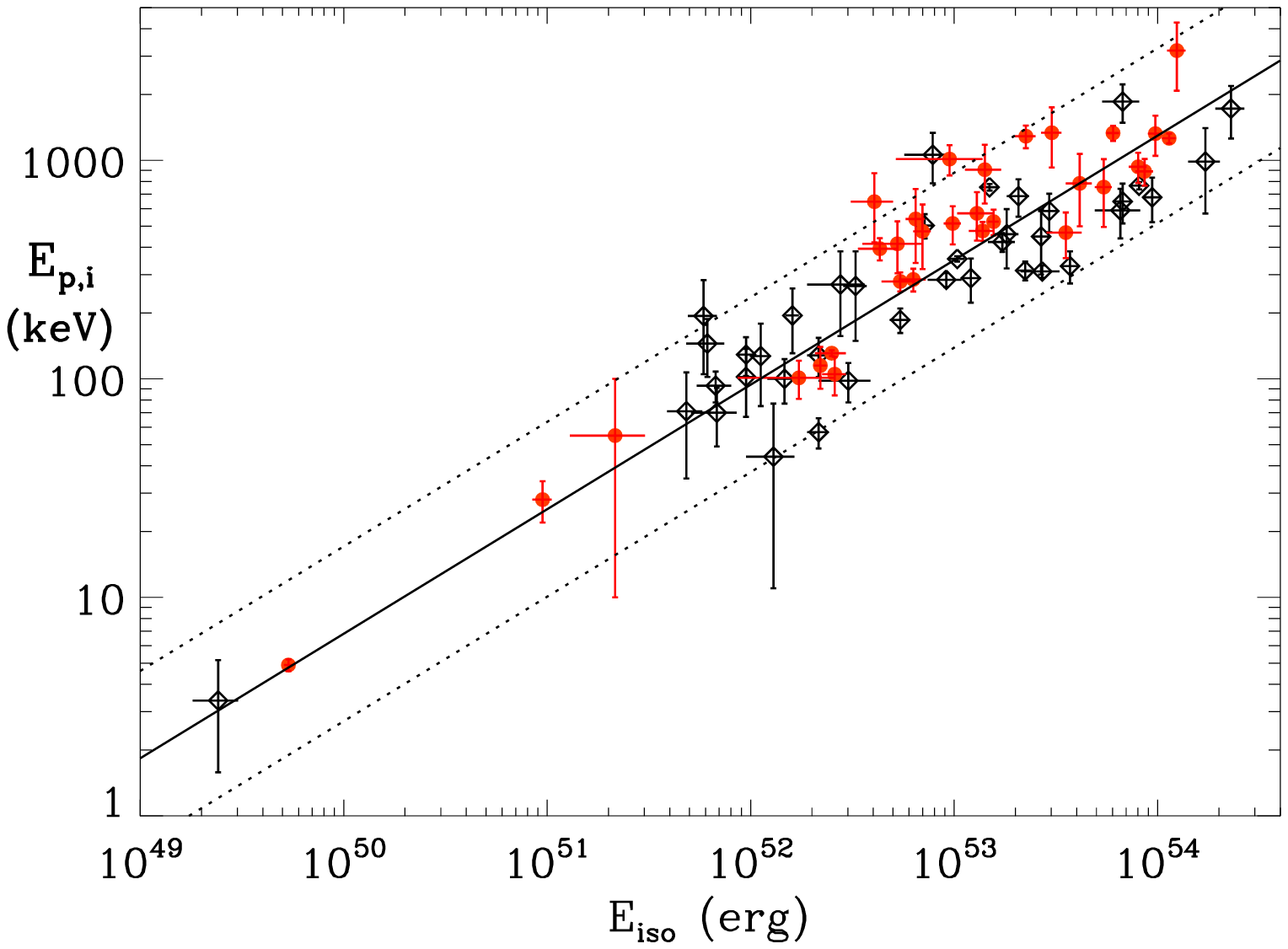},\includegraphics[width=8.8cm]{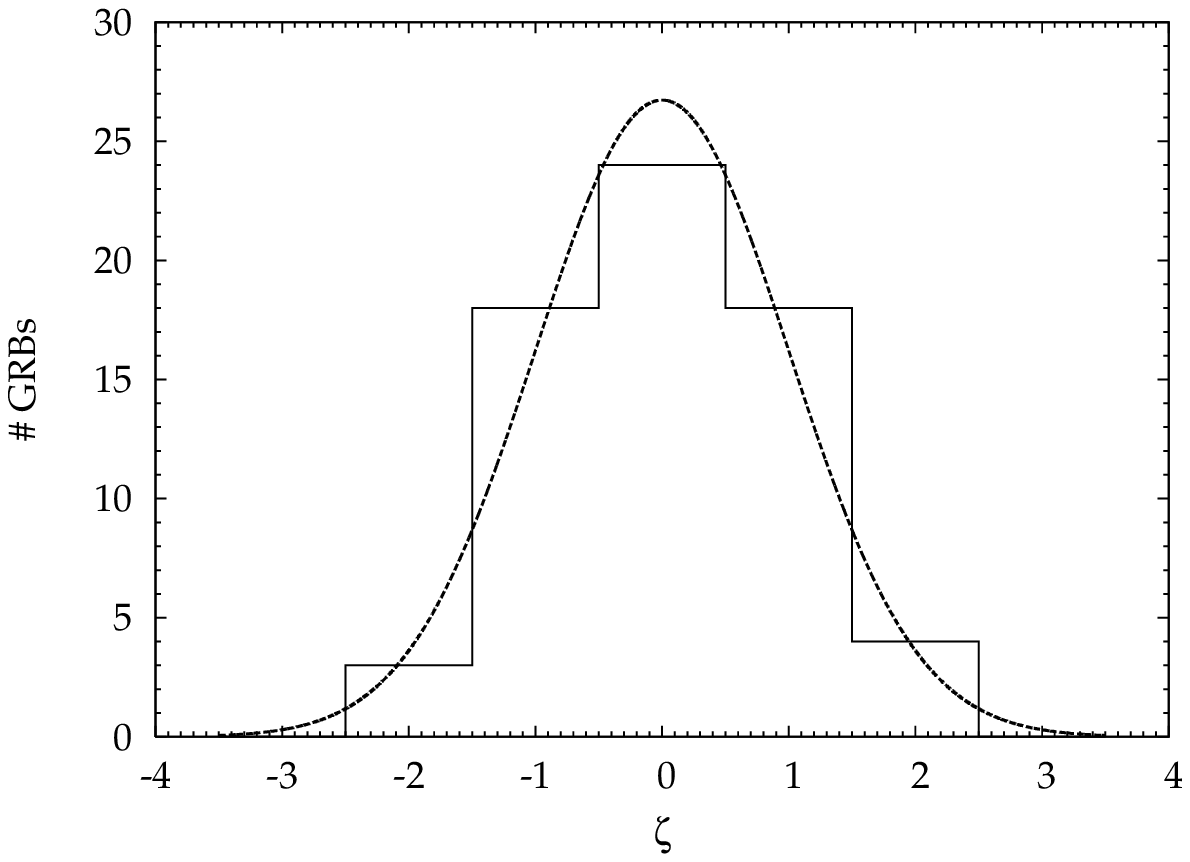}} 
\caption{Left: location in the \epeiso{} plane of the 70 GRBs and XRFs with firm 
estimates of redshift and \epo included in our sample (\eiso{} computed following 
Amati (2006) and assuming 
a cosmology with H$_{\rm 0} = 70$ km s$^{-1}$ Mpc$^{-1}$, $\Omega_{\rm M} = 0.3$ and 
$\Omega_{\Lambda} = 0.7$). Red dots are \swift{} GRBs. Black dots are GRBs discovered by other satellites. 
The best--fit  power--law is the continuous line ($\pm$2\sext{} region). Right: distribution of 
the normalised scatter (see, e.g., Rossi et al. 2008 for definition and method) 
of the \epeiso{} correlation for H$_{\rm 0} = 70$ km s$^{-1}$ Mpc$^{-1}$, $\Omega_{\rm M} = 0.3$ and
$\Omega_{\Lambda} = 0.7$); the normalised Gaussian 
is superimposed 
to the data.} 
\end{figure*}

\section{The updated \epeiso{} sample}

Previous analyses of the \epeiso{} plane of GRBs shows that
different classes of GRBs exhibit different behaviours: while normal
long GRBs and X--Ray Flashes (XRF, i.e. particularly soft bursts)
follow the \epeiso{} correlation, short GRBs and the peculiar very close
and sub--energetic GRB\,980425 do not (Amati et al. 2007).
Therefore, in our analysis we considered only long GRB/XRF. The sample of long GRB/XRF
used in this paper (updated to April, 11 2008) is reported in Table 1 and
includes 70 GRBs. We report, for each GRB, the redshift, the
fluence and the cosmological
rest--frame spectral peak energy, \epi{}=\epo{}$\times$(1+$z$). 
The redshift distribution covers a
broad range of $z$, from 0.033 to 6.3, thus extending far beyond that
of Type Ia SNe (z $<$$\sim$1.7).  
For \swift{} GRBs, the \epi{} value derived from BAT spectral
analysis alone were conservatively taken from the results reported
by the BAT team 
\citep{Sakamoto08a,Sakamoto08b}.  Other BAT \epi{} values
reported in the literature were not considered, because
either they were not confirmed by \citet{Sakamoto08a,Sakamoto08b} refined
analysis (e.g., Cabrera et al. 2007\nocite{Cabrera07}) 
or they are based on speculative methods
\citep{Butler07}.

In Table 1 we also report the values of \eiso{} computed
from published spectral parameters and fluences by following the
standard method described, e.g., by \citet{Amati06} and assuming H$_{\rm
0} = 70$ km s$^{-1}$ Mpc$^{-1}$, $\Omega_{\rm M} = 0.3$ and
$\Omega_{\Lambda} = 0.7$ (references are reported in Table 1). 
An analysis of Figure 1 (left panel) shows that, for this "standard"
cosmology,  all GRBs in our
sample follow the \epeiso{} correlation; in particular, \swift{} GRBs are
well consistent with the \epeiso{} correlation based on GRBs discovered
by other instruments characterised by different
trigger thresholds and sensitivities (as a function of energy). This
fact points out that the \epeiso{} correlation is not affected by
significant selection effects (see also Ghirlanda et al. 2008\nocite{Ghirlanda08}).

If we fit the
data of Figure 1 with a simple power--law, we find an index $m =0.57\pm
0.01$ and a normalization $K=94\pm2$, consistent with results of
previous analyses (e.g., Amati 2006). However, despite the very
high significance of the correlation (Spearman's $\rho=0.872$ for 70
events), the fit with a power--law provides an highly unacceptable
\cqr value (408/68). This is clear evidence of the
existence of a significant extrinsic scatter, which is
superimposed to the Poissonian one and 
implies the existence of "hidden" parameters,
connected with GRB phenomenology, 
which contribute to define the location of a GRB in the \epeiso{} plane. 
It should be noted that systematics affecting both \epi{} and \eiso{}
may contribute to the extrinsic scatter, even if there is evidence that they
are not a dominant component \citep{Landi06}. 

\begin{figure*} 
\centerline{\includegraphics[width=9.5cm]{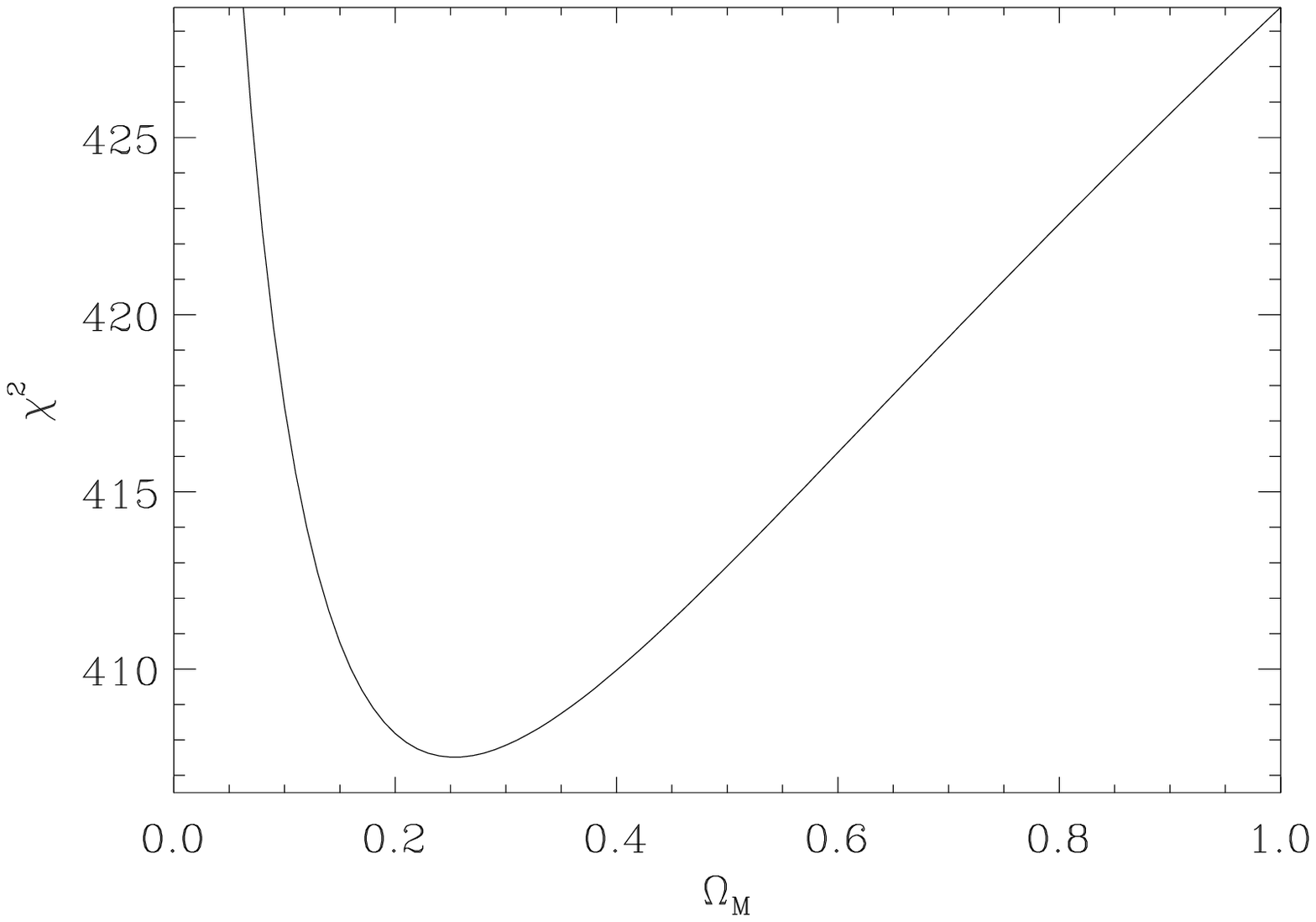}\includegraphics[width=9.5cm]{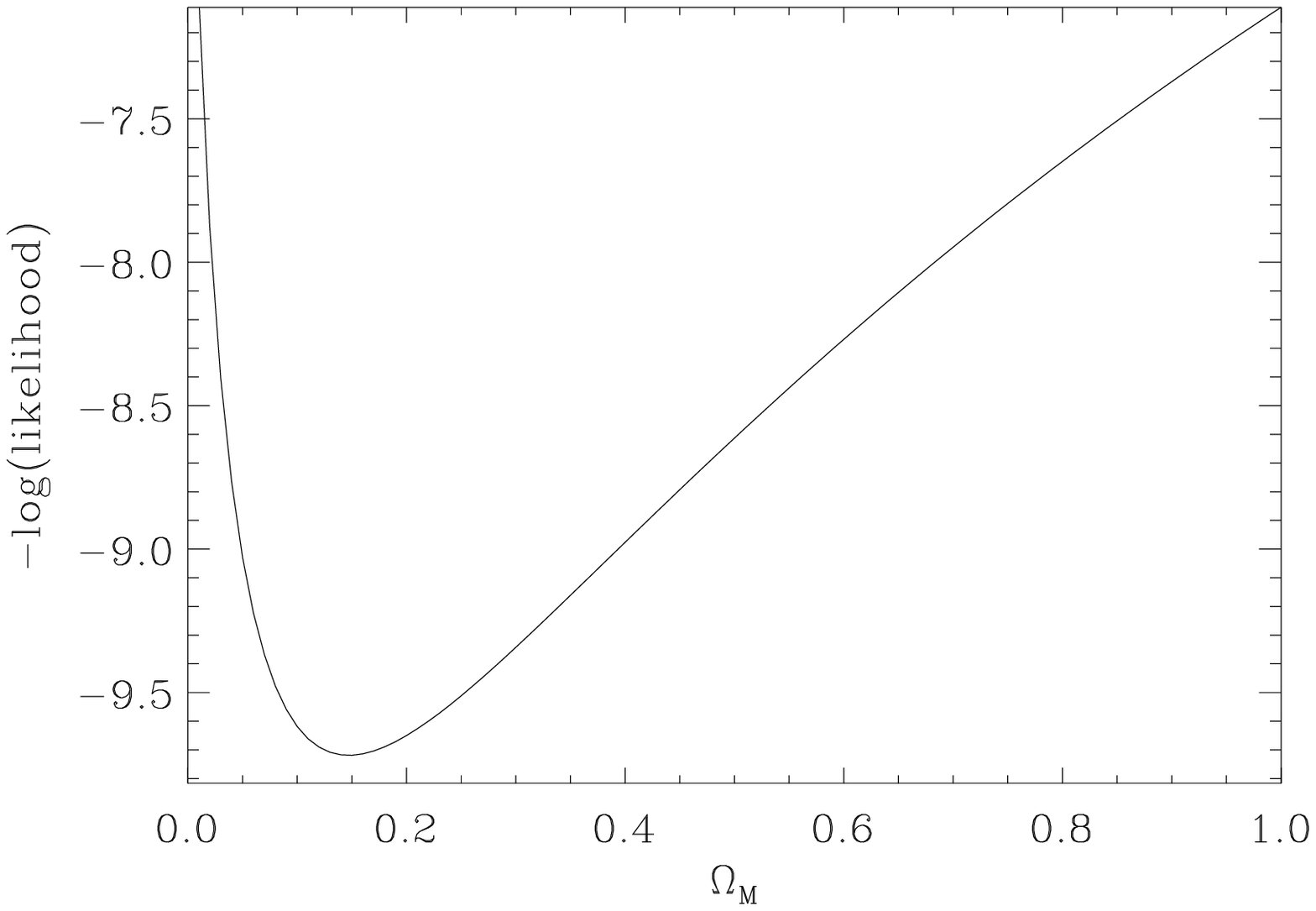}} 
\caption{Value of the $\chisq${} of the simple power--law fit (left) and of the 
\mloglik{} of the fit with the MLM method which accounts for extrinsic variance 
(right) 
as a 
function of \omegam{} in the assumption of a flat universe. 
} 
\end{figure*}
\begin{figure*}
\centerline{\includegraphics[width=9.5cm]{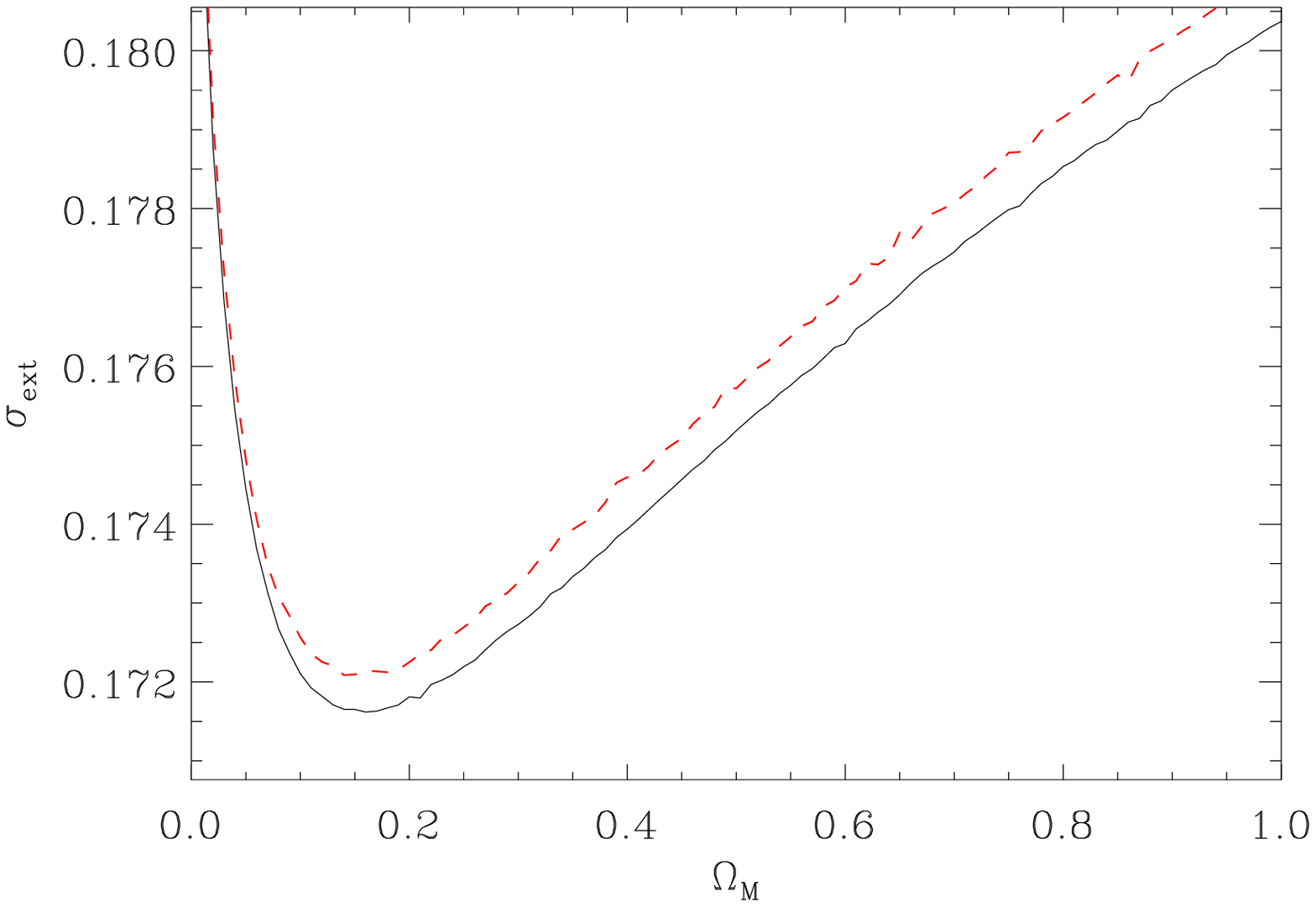}\includegraphics[width=9.5cm]{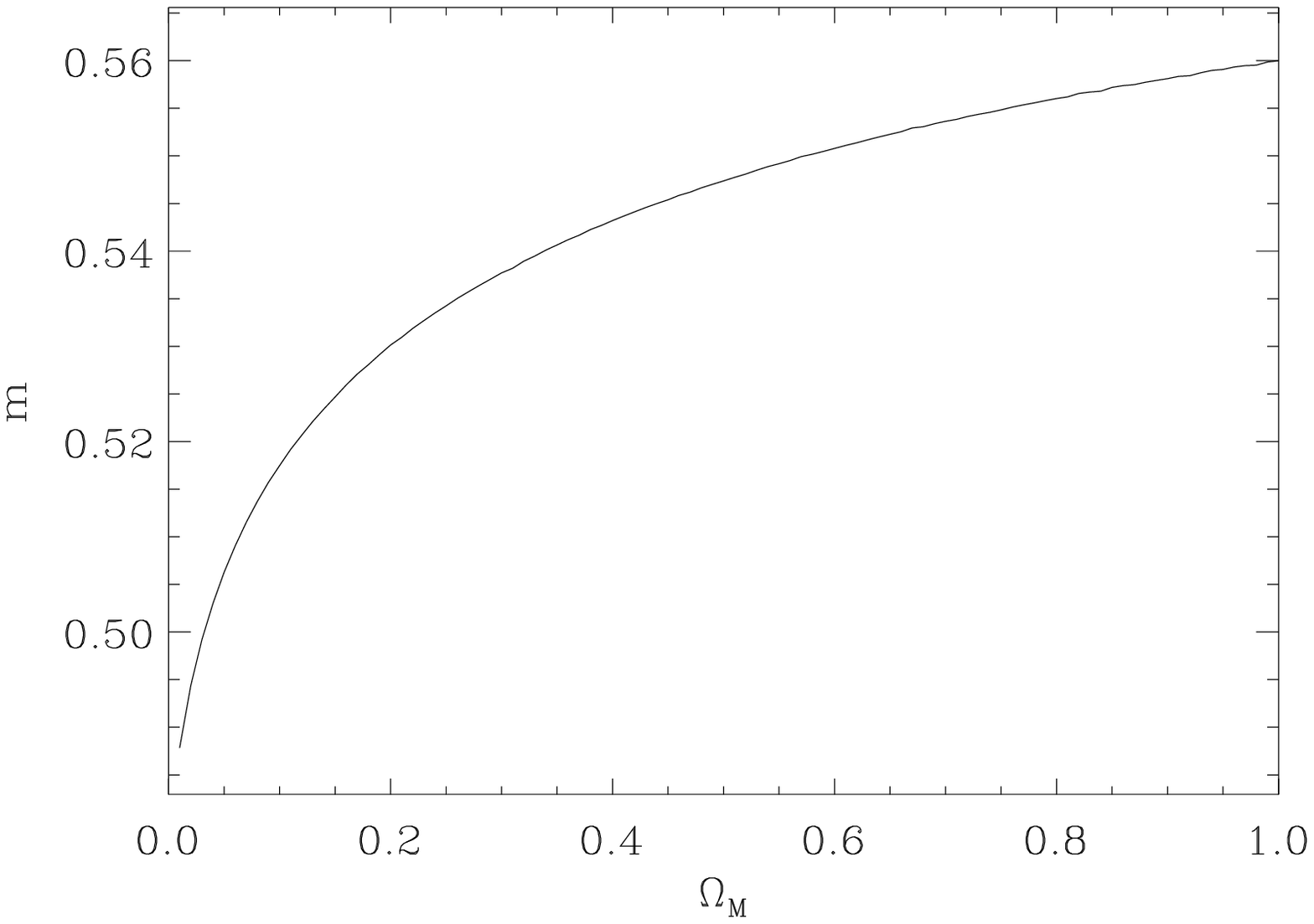}}
\caption{Value of the extrinsic scatter \sext{} (left) and power--law index
$m$ (right) 
as a function of
\omegam{}, obtained by fitting the correlation with the MLM method which accounts for extrinsic variance,
for a flat universe. In the left panel,
we also show, as a dashed line, the result obtained
by using a different \loglik{} (see text).}
\end{figure*}

\section{Cosmological parameters derived from the \epeiso{} correlation}

The main goal of this paper is to investigate if the extrinsic
dispersion of the \epeiso{} correlation is sensitive at varying the
values of the cosmological parameters \omegam{} and \omegal.  We
emphasize that this method does not suffer from circularity, since we
do not assume an \epeiso{} relation based on a particular choice of
the cosmological parameters or calibrate it by using other
cosmological probes. 

As a first step, under the assumption of a flat universe, we studied the
trend of the $\chisq${} values obtained by fitting the \epeiso{} 
correlation with a simple power--law, for different choices of
\omegam{}. 
Here and in the the following, we assumed \h0 = 70 km s$^{-1}$ Mpc$^{-1}$.
Figure 2 (left) shows that the extrinsic
scatter of the correlation indeed decreases with cosmology, and minimizes
for \omegam{} $\sim$0.25. 
This result is in qualitative agreement with
the one obtained, e.g., with SNe \citep{Perlmutter98,Perlmutter99, Riess98,
Riess04,Astier06}.

In order to better characterize the dependence of the \epeiso{} extrinsic 
scatter on cosmology, we have used the
maximum likelihood method (hereafter MLM) as discussed  by
\citet{Dagostini05} (see also Guidorzi et al. 2006, Amati 2006 and Rossi et
al. 2008 for other applications of this methodology). This method
assumes that, if the correlated data (x$_i$,y$_i$) can be described by 
a linear function $Y=mX+q$ with the addition of an extrinsic scatter \sext{}
among the free parameters, the optimal values
of the parameters ($m$, $q$, and \sext{}) can be obtained by minimizing the
\mloglik{} function, in which the uncertainties on both (x$_i$,y$_i$) 
are taken into account.
The general \loglik{} function  is given by:
\begin{eqnarray}
\displaystyle \log{p(m,q,\sigma_x,\sigma_y|\{x_i,y_i,\sigma_{x,i},\sigma_{y,i}\})}
=~~~~~~~~~~~~~~~~~~~~~~~\nonumber\\
~~~~\frac{1}{2}\,\sum_{i=1}^N \Big[\log{\Big(\frac{1}{2 \pi (\sigma_y^2 + m^2\,\sigma_{x}^2 + \sigma_{y,i}^2 + m^2\,\sigma_{x,i}^2)}\Big)}\quad +\nonumber\\
~~~~-\quad\frac{(y_i - m\,x_i - q)^2}{\sigma_y^2 + m^2\,\sigma_{x}^2 + \sigma_{y,i}^2 + m^2\,\sigma_{x,i}^2}\Big]
\label{eq:prior}
\end{eqnarray}
where, in our case,
{\rm log}(\epi), $\sigma$$_x$ = 0 and $\sigma$$_y$ = \sext. We remark that setting
$\sigma$$_x$ = 0 does not affect the general validity of this formula.

As an example, the fit of the correlation with this method by using the
\eiso{} values reported in Table 1
provides the following
parameters value: $m =0.54\pm0.03$, $K=98\pm7$ and $\sigma_{\rm
ext}=0.17\pm0.02$ (68\% c.l.; consistent with the results reported by Amati 2006
for a smaller sample of 41 GRBs). 
In Figure 1 (right) we show the distribution of the
normalised scatter of the data (see, e.g., Rossi et al. 2008 for 
details on the computation of this quantity) with superimposed the 
normalised Gaussian.
It is apparent that the
scatter of the \epeiso{} correlation is indeed dominated by the
extrinsic (i.e. non--Poissonian) variance.

For the goals of this paper, we repeated the above analysis 
by varying \omegam, always under
the assumption of a flat universe.  Figure 2 (right) and Figure 3
(left) show that the values of the \mloglik{} and of the extrinsic
variance \sext{} are indeed sensitive to \omegam, both showing a clear
minimum around \omegam{} $\sim$0.15.  Also the slope of the
correlation is sensitive to the adopted cosmology, as shown in Figure
3 (right). By using the probability density function, the MLM also
allows us to constrain
\omegam{}, that results to be in the range 0.04--0.40 at 68\%
and in the range 0.02--0.68 at 90\% c.l (Table 2). An \omegam{} value of 1 can be
exluded at $\sim$99.9\% c.l.  

If we release the flat universe hypothesis and let \omegam{} and
\omegal{} vary independently (Figure 4, left), we still find evidence
for an universe with a low value of \omegam{} (0.04--0.50 at 68\%
c.l.). Only an upper limit of $\sim$1.05 can be set to \omegal{} (see
Table 2). This fact is not surprising, given that the redshift
distribution of GRBs is expected to produce very vertically elongated
contours in the \omegam{} -- \omegal{} plane
\citep{Ghirlanda06,Kodama08}. 
The complementarity of GRB with other cosmological probes can be seen
in Figure 4, where we plot, in addition to our results, the contours 
derived from SN Ia by \citet{Astier06}. As can be seen, already with
the present sample of GRB, significantly improved  constraints on
\omegam{} can be obtained by the combination of the two probes.

We have also investigated the estimate of the free paramaters of the
the inverse relation $X=mY+q$ plus the extrinsic scatter parameter 
$\sigma_{ext} = \sigma_x$. The results are fully consistent with previous ones
and satisfy the expectation that $\sqrt{mm'}= r_{xy}$, where $r_{xy}$ is the Pearson's 
weighted linear-correlation coefficient (e.g., Bevington 1969, Bendat \& Piersol
2000\nocite{Bevington69,Bendat00}).
For a flat universe with \omegam = 0.3, we find $m = 0.54$, $m' = 1.57$, $\sqrt{mm'} = 0.92$ and
$r_{xy} = 0.92$. 

Finally, we tested our results using the likelihood function suggested by
\citet{Reichart01}. 
The results obtained
are fully consistent with the results above (Figure 2, left). 
This shows that the constraints that we have derived on
the cosmological parameters do not depend significantly on the adopted
statistical methodology and confirms the general soundness of our
approach.

\begin{table}
\begin{minipage}{\columnwidth}
\caption{68\% and 90\% c.l. ranges of \omegam{} (both by assuming a flat universe or
by letting \omegam{} and \omegal{} to vary independently)
as determined with different methods
applied to the present sample of 70 GRBs.
}
\begin{tabular}{lcccc}
\hline
 Method  & c.l. & \omegam (flat) & \omegam & 
\omegal \\ 
\hline
scatter &  68\% & 0.04 -- 0.40 &  0.04 -- 0.50 & $<$1.05 \\ 
 &  90\% & 0.02 -- 0.68 &  0.01 -- 0.75 & $<$1.15 \\
scatter  & 68\% & 0.04 -- 0.40 & 0.05 -- 0.41 & $<$1.05 \\
(self calib.)  & 90\% & 0.02 -- 0.67 & 0.01 -- 0.73 & $<$1.13 \\
scatter &  68\% & 0.03 -- 0.28 & 0.03 -- 0.33 & $<$1.10 \\
($m = 0.5$)  & 90\% & 0.01 -- 0.49 & 0.01 -- 0.53 & $<$1.17 \\
\hline
\end{tabular}
\end{minipage}
\end{table}

\begin{figure*}
\centerline{\includegraphics[width=9.5cm]{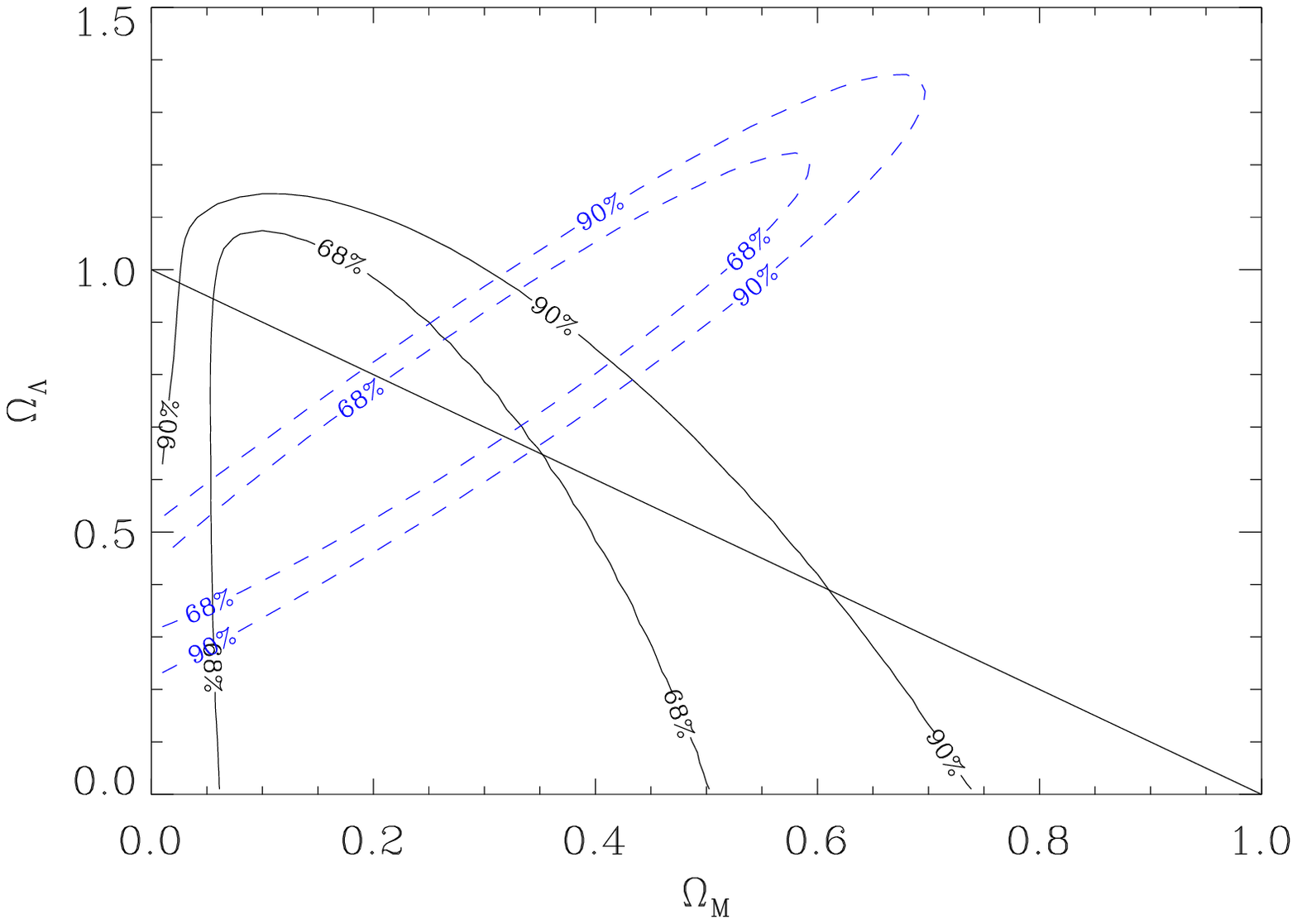}\includegraphics[width=9.5cm]{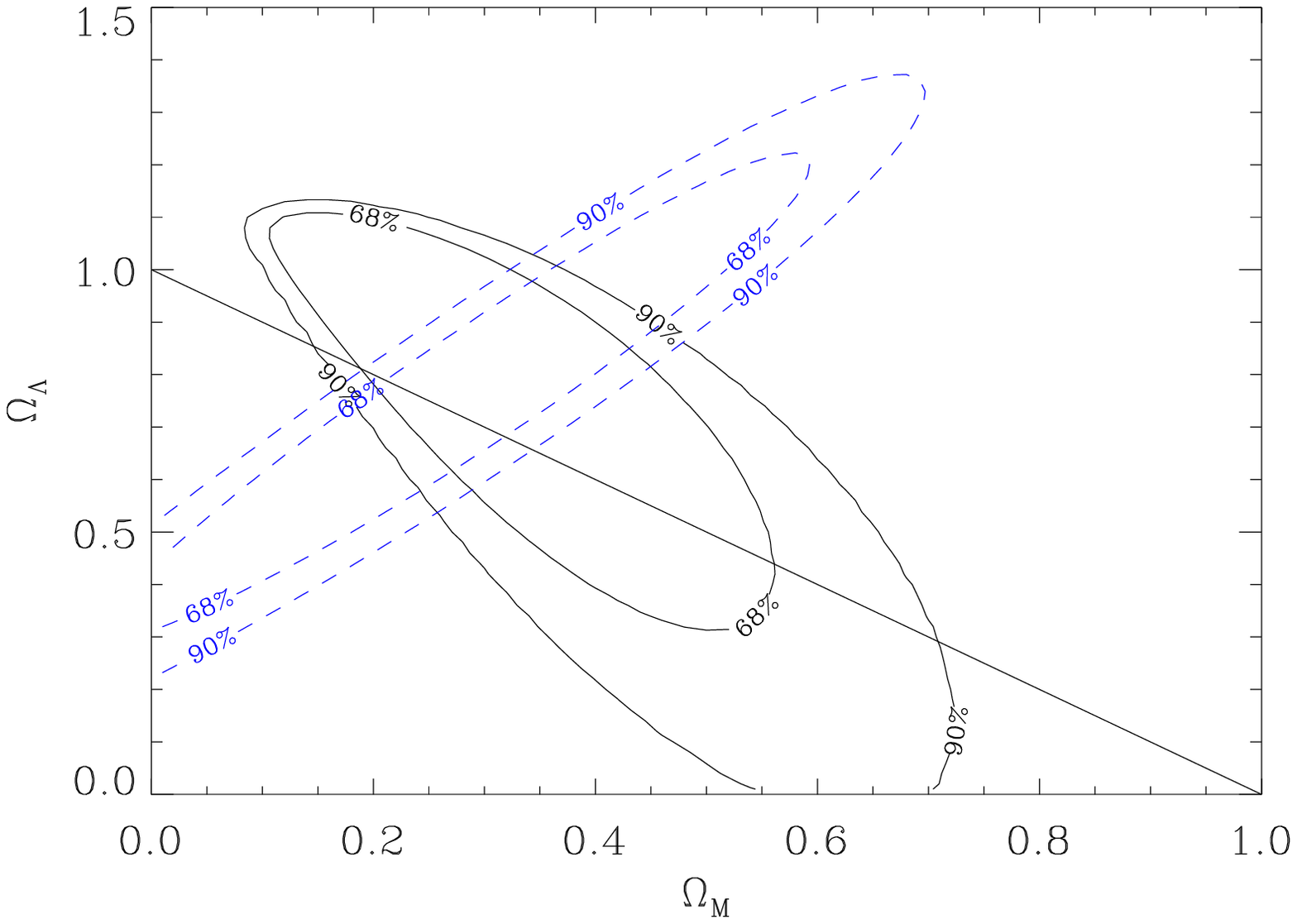}}
\caption{Contour confidence levels of
\omegam{} and \omegal{}, 
obtained by fitting the correlation with the MLM method which accounts for
 extrinsic variance,
with the present sample of 70 GRBs (left) and with the 
improved sample of 70 $+$ 150 GRBs expected from next experiments (right; see text). 
In both panels we also show as blue dashed contours the constraints derived 
from SN Ia by Astier et al. (2006).
}
\end{figure*}
\begin{figure*}
\centerline{\includegraphics[width=9.5cm]{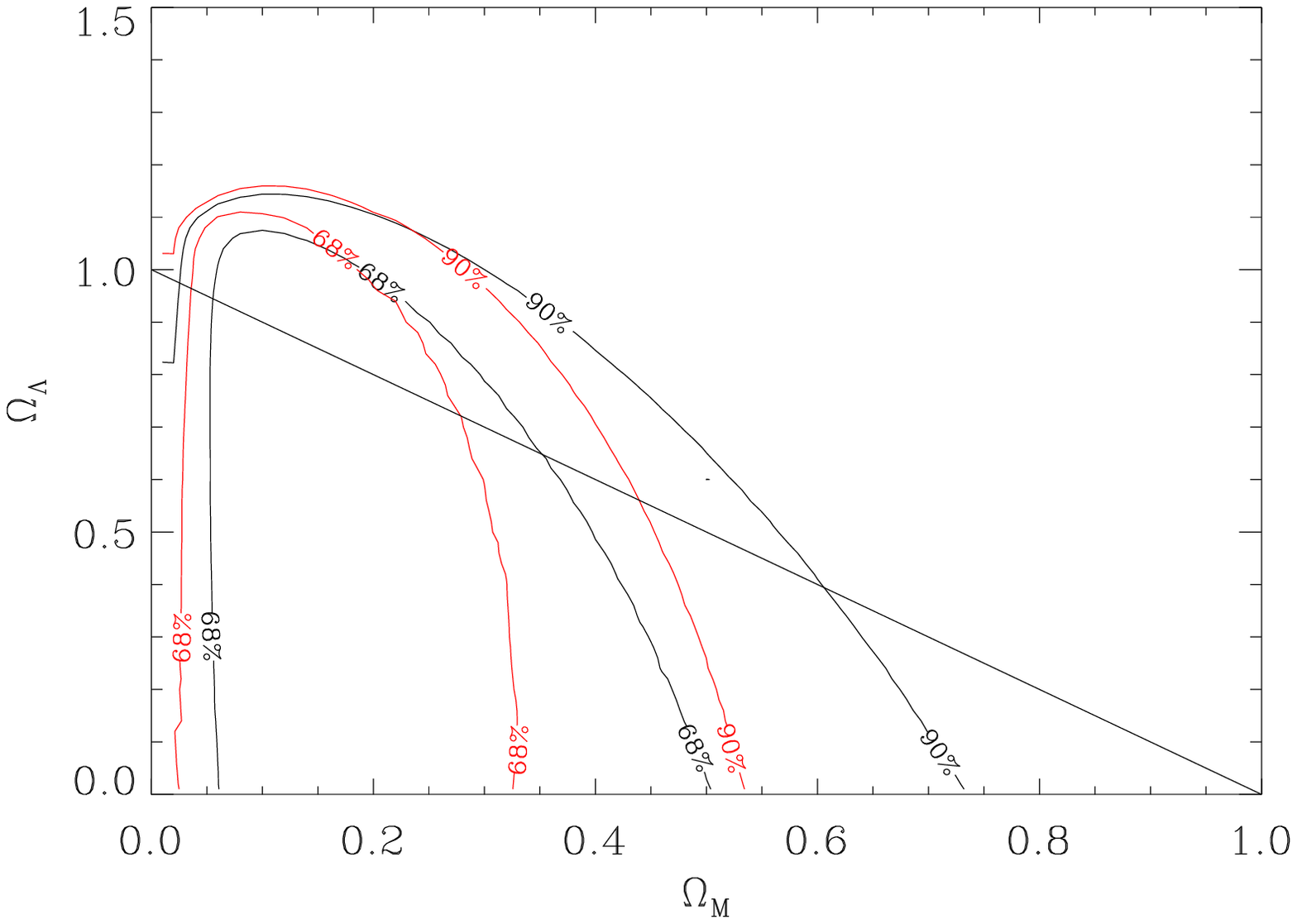}\includegraphics[width=9.5cm]{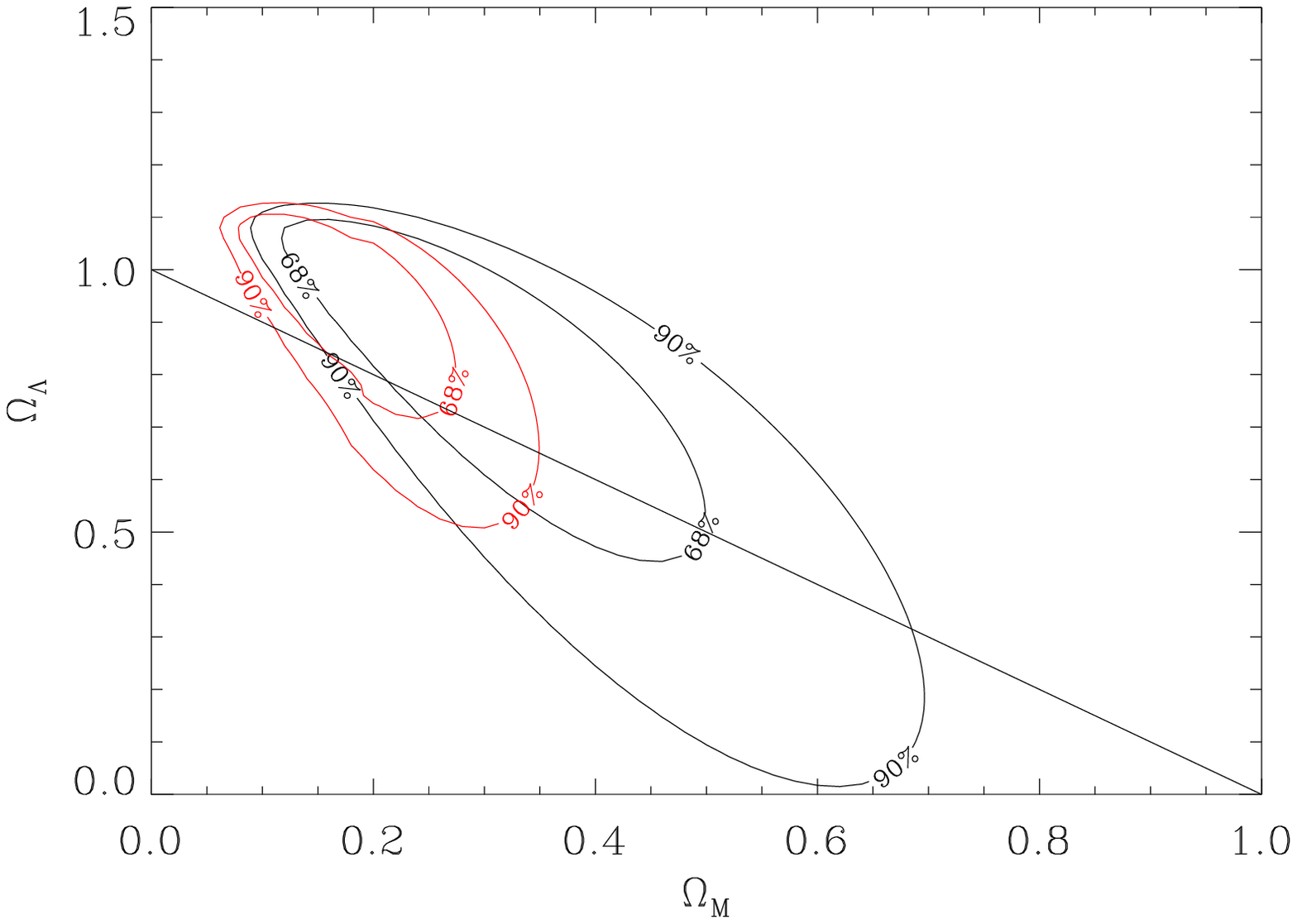}}
\caption{Contour confidence levels of
\omegam{} and \omegal{} 
obtained by fitting the correlation with MLM method which accounts for
 extrinsic variance
with $m$ calibrated based on GRBs lying at $0.8<z<1.2$ (black contours) and
$m$ fixed at 0.5 (red contours); see text.
Left: present sample of 70 GRB; right: 
improved sample of present 70 GRBs $+$ 150 GRBs expected from next experiments. 
}
\end{figure*}

\begin{figure*}
\centerline{\includegraphics[width=9.5cm]{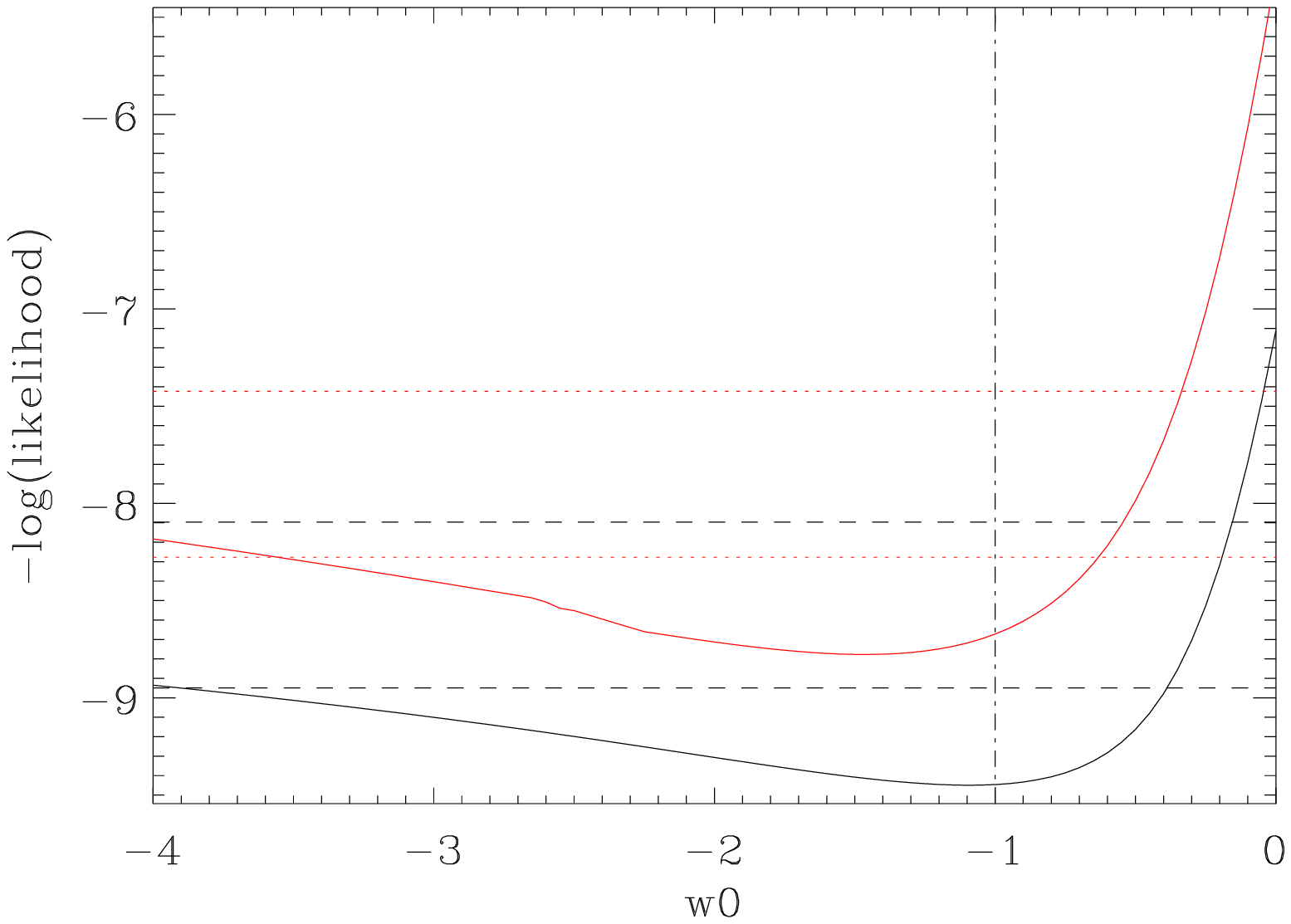}\includegraphics[width=9.5cm]{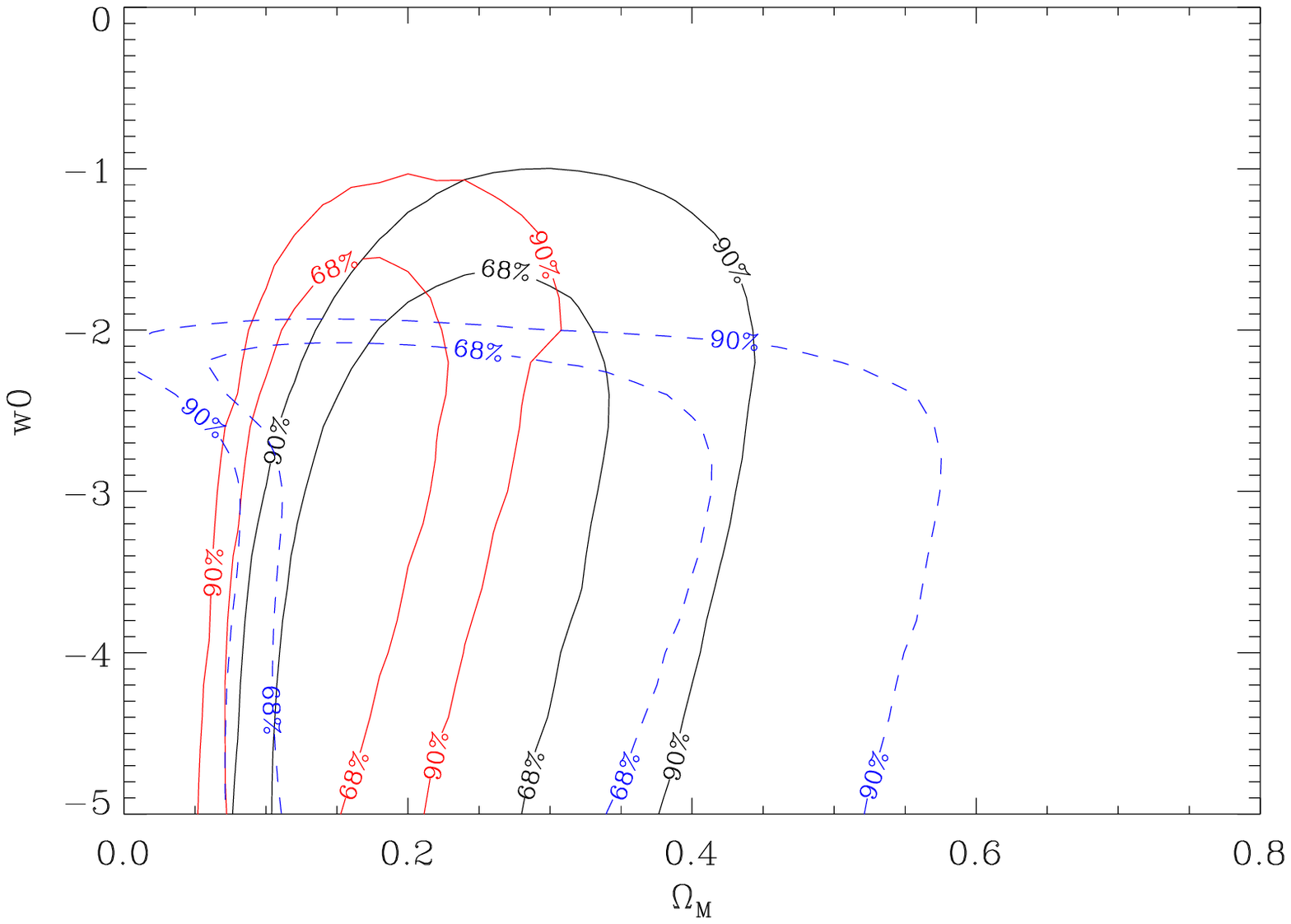}}
\caption{Left: Value of the \mloglik{}
as a function of
the parameter of the dark energy equation of state $w_0$ (see text)
in the assumption of a flat universe with 
\omegam{} = 0.27 and \omegal{} = 0.73 obtained with the 
present sample of 70 GRB. Right: contour confidence levels of \omegam{} and $w_0$
in the assumption of a flat universe 
obtained with the improved sample of present 70 GRBs $+$ 150 GRBs expected from next experiments
(continuous line: $w_a$ = 0; 
dashed lines: $w_a$=4).
In both figures, the red lines indicate the results obtained by fixing $m$ at 0.5 .
}
\end{figure*}

\section{Future perspectives}

\subsection{Enrichment and redshift extension of the sample}

In order to estimate the accuracy on cosmological parameters
achievable by ongoing and future experiments, we have carried out a
number of simulations based on an enriched sample of GRBs.  By using
the slope, normalization and extrinsic scatter of the \epeiso{}
correlation obtained with the real sample of 70 GRBs and assuming, as
an example, \omegam=0.27, \omegal=0.73, we generated 150 random GRBs
with an accuracy on \epi{} and \eiso{} of 20\%. 
For 90\% of the simulated GRBs, the redshift was randomly chosen from
the GRB redshift distribution of our sample. For the remaining 10\%
of simulated GRBs we assumed a value of redshift larger than 6. 
The number of GRBs, and the accuracy on \epi{} and \eiso{}, of the
simulated sample are those that will realistically be provided in the
next years by joint observations of \swift{} and GLAST/GBM (plus
also Konus-Wind, Suzaku, RHESSI, AGILE, etc.).
Also the extension of the redshift distribution up to at
least $z=9$ is consistent both with theoretical predictions based on
star formation rate evolution and the expected improvement in the
sensitivity of hard X--ray and optical/near--infrared telescopes, that
will be used to detect and follow--up GRBs and measure GRB redshift
(e.g., Salvaterra et al.  2007\nocite{Salvaterra07} Salvaterra \& Chincarini 2007).

In Figure 4 (right) we show the confidence level contours for \omegam
and \omegal{} obtained by applying the method described in Section 3 to
the sample made of the 70 ``real'' GRBs plus the 150 ``simulated''
ones. In Table 3, we report the constraints on cosmological
parameters both in the assumption of a flat universe and by varying
simultaneously \omegam{} and \omegal. A simple comparison with
Table 2 shows that the simulated sample decreases the
uncertainty range of \omegam{} from 10 to only a factor 2. 
In addition, as can be seen in Figure 4 (right), significantly
improved constraints on both \omegam{} and \omegal{} will
be obtained by the 
combination of the contours
obtained from GRBs with those from other probes, like, e.g., SN Ia.

Finally we note that the \epega{} correlation, because it needs the
measure of a third observable (i.e. the break time in the optical
afterglow light curve) requires the estimates of $z$ and \epi{} for a
number of GRBs larger by more than a factor of 3 than required by the
\epeiso{} correlation. For instance, the simulations performed by
\citet{Ghirlanda06} with 150 GRBs correspond to a ``real'' sample of
at least $\sim$500 GRBs having both $z$ and \epi{} measured.

\subsection{Calibrating the \epeiso{} correlation}

One of the main problems with the use of GRB correlations for
cosmology is the lack of low--redshift GRBs (i.e. lying at z $<<$1)
allowing for a cosmology--independent calibration similarly to type Ia
SNe. This problem can be partly overcome by fitting the correlation in
a sub--sample of GRBs lying at similar redshifts. This technique
was proposed for future calibration of the slope of
the \epega{} correlation (e.g., Ghirlanda et al. 2006), 
but can be effectively used for the \epeiso{}
correlation, due to the much larger number of events that can be
included in the \epeiso{} sample with respect to the \epega{} one. By
studying the redshift distribution of the GRBs in our sample, we find
that there are 18 events lying at redshift between $\sim$0.8 and
1.2. The analysis of this sub--sample with the MLM
method provides, indeed, evidence of a very low variation of $m$ as a
function of cosmological parameters. For instance, by assuming a flat
universe, we find that the $m$ best--fit values are in the range
$m$=0.493--0.496, to be compared to $m$=0.477--0.560 obtained on the
whole sample of 70 GRB. The value of $m$ holds stable within
this interval even by varying both \omegam{} and \omegal{} in the [0,1]
range.  The dependence of $m$ on the dark energy equation of state
(parametrised as discussed in next section) is also within this
interval. However, when taking into account the uncertainty
resulting from the fits, the constraint on the calibrated slope of the
\epeiso{} correlation becomes larger: $m$= 0.44--0.55, for both a flat
and general universe. 
Thus, the applicability of the method is currently hampered by the
large uncertainty affecting $m$, which results in a marginal
improvements of the c.l. of cosmological parameters (Figure 5, left;
Table 2). 
On the other hand, if
we apply this calibration method to the simulated sample of 70 +
150 GRB, the improvement in the estimates of cosmological parameters
is significant (see Table 3 and compare Figure 5, right, with Figure
4, right). 

A second possibility to calibrate the slope of the correlation is
to derive it on robust physical basis. Several analyses reported
above or elsewhere (e.g., Ghirlanda et al. 2008) point out a slope of
$m$=0.5.  This value is also predicted by several theoretical models,
involving in various forms synchrotron, inverse Compton, thermal and
Comptonised thermal emissions; see, e.g., \citet{Zhang02,Thompson07}.  
Figure 5 (red contours) and Table 3 illustrate quite clearly
that the correlation fitted with the MLM method after
assuming $m=0.5$ improves significantly the accuracy degree of the 
cosmological parameter measurements.

\subsection{Investigating the evolution of the dark energy}

The GRB redshift distribution extends up to $z >\sim 6$, which is
well above that of SNe-Ia ($z\sim 1.7$).  Therefore, at least in
principle, GRBs are powerful tools to study the evolution of dark
energy with the redshift. In the following, we adopt the
parametrization of the dark energy equation of state proposed by
\citet{Chevallier01,Linder05}, i.e.  $P=w(z)\rho$, where:
\begin{equation}
w(z)= w_{0} + {w_{a}z \over 1+z}
\label{param}
\end{equation}

As shown in Figure 6 (left), with our sample of 70 GRBs we find that the
\mloglik{} computed with the MLM by assuming
a flat universe with \omegam{} = 0.27 and by setting $w_a$ = 0 is
indeed sensitive to $w_0$, with a minimum around $-1$ (corresponding
to a cosmological constant).  Unfortunately, within the limits of
present GRB sample no significant constraints on $w_0$ can be
provided. Even after assuming $m=0.5$,
$w_0$ can be constrained only to be between $-$3.7 and
$-$0.7 at 68\% c.l. In Figure 6 (right) we show the \omegam{} -- $w_0$
confidence level contours obtained with the sample of 70 real +
150 simulated GRBs by assuming a flat universe with
$w_a$ = 0 (continuous lines) and $w_a$ = 4 (dashed line). 
As can be seen, such a sample would provide clear evidence 
of $w_0$$<$$-$1 and some hints on $w_a$.

\begin{table}
\begin{minipage}{\columnwidth}
\caption{68\% and 90\% c.l. ranges of \omegam{} (both by assuming a flat universe or
by letting \omegam{} and \omegal{} to vary independently)
as determined with different methods 
applied to a sample composed by the present sample of 70 GRBs plus 150 simulated
GRBs with known $z$ and \epi{} expected to be provided in the near future 
by present and planned GRB experiments (see
text).
}
\begin{center}
\begin{tabular}{llcccc}
\hline
 Method  &  c.l. & \omegam (flat)  & \omegam  & 
\omegal \\ 
\hline
scatter & 68\% & 0.21 -- 0.50 & 
0.11 -- 0.57 & 0.35 -- 1.12 \\
 &  90\% &  0.16 -- 0.66 & 
0.08 -- 0.73 & $<$1.15 \\
 scatter & 68\% & 0.22 -- 0.48 &
0.12 -- 0.49 & 0.47 - 1.10 \\
 (self calib.)  & 90\% & 0.17 -- 0.59 &
0.10 -- 0.69 & 0.10 - 1.11 \\
 scatter &  68\% & 0.13 -- 0.27 &
0.18 -- 0.28 & 0.62 -- 1.10 \\
 ($m = 0.5$)  & 90\% & 0.11 -- 0.35 &
0.06 -- 0.35 & 0.50 -- 1.12 \\
\hline
\end{tabular}
\end{center}
\end{minipage}
\end{table}

\section{Conclusions}

We have used an updated sample of 70 GRBs aimed at deriving the
cosmological parameters \omegam{} and \omegal{} from the \epeiso{}
correlation of GRBs. With respect to previous attempts mainly based on
the \epega{} and \epeiso--T$_{0.45}$ correlations, the use of
\epeiso{} correlation has the advantages of requiring only two
parameters, both directly inferred from observations. This fact allows
for the use of a richer sample of GRBs (e.g., a factor 3-4 larger
than used in the \epega correlation), and a reduction of systematics,
with respect to three--parameters spectrum--energy correlations.

Our method consists of studying the dependence of the dispersion of
the \epeiso{} correlation on cosmological parameters by adopting a
maximum likelihood method, which allowed us to parameterise
and quantify correctly the extrinsic (i.e. non--Poissonian) scatter
component. From our analysis, a number of significant results do
emerge:

a) a simple power--law fit shows a clear trend of the dispersion, with
a minimum at \omegam{} = 0.2--0.4 (for a flat universe). By refining the
fit with the MLM, we found that both the extrinsic
variance and the slope of the correlation show a significant
dependence on cosmology. From the study of the \loglik{} we derived
\omegam{} = 0.04--0.40 at 68\% c.l and 0.02-0.68\% at 90\% confidence
levels;

b) after assuming ``a priori'' a slope $m=0.5$, consistently with both
observations and the predictions of several models for GRB prompt
emissions, we derive \omegam{} = 0.01--0.49 at 90\%;

c) if we release the assumption of a flat universe, we still find
evidence for a value of \omegam{} $<1$ (0.04--0.50 at 68\% c.l.) and a weak
dependence of the dispersion of the \epeiso{} correlation on
$\Omega_{\Lambda}$ with an upper limit of 1.15 (90\% c.l.);

d) our study does not make assumptions on the \epeiso{} correlation or
make use of independent calibrators to set the ``zero point'' of the
relation, therefore our approach does not suffer from circularity and
provides independent evidence for the existence of a gravitationally
repulsive energy component ("dark energy") which accounts
for a large fraction of the energy density of the universe;

e) in a flat universe, the \epeiso{} correlation is also (weakly)
sensitive to the $w_0$ parameter of the equation of state of dark
energy;

f) we have simulated the impact of ongoing/planned GRB experiments
(e.g., \swift{} + GLAST) on the future \omegam{} and
\omegal{} measurements and shown that the uncertainties can be decreased by almost
an order of magnitude with respect to those obtained with the current
GRB sample.


\label{lastpage}

\end{document}